\documentclass[aps,pre,showpacs,superscriptaddress,longbibliography,twocolumn,floatfix]{revtex4-2}
\usepackage{graphicx}
\usepackage{amsmath}
\usepackage{amssymb}
\usepackage{bm}
\usepackage{times}
\usepackage[pdftex]{hyperref}
\hypersetup{colorlinks=true,linkcolor=blue,citecolor=blue,urlcolor=blue}
\usepackage{verbatim}
\usepackage{microtype}
\usepackage{subcaption} 
\usepackage{tikz}
\usetikzlibrary{spy}
\usepackage[justification=raggedright,singlelinecheck=false]{caption}

\begin{document}

\preprint{APS/123-QED}

\title{Dislocations, vacancies and interstitials in the two-dimensional one-component plasma}

\author{G. Vilella Nilsson}
\affiliation{Department of Physics, Lund University, SE-221 00 Lund, Sweden}

\author{M. A. Moore}
\affiliation{Department of Physics and Astronomy, University of Manchester, Manchester M13 9PL, United Kingdom}
\date{\today}
\begin{abstract}

The energetics and stability of dislocations, vacancies and, interstitials in the one-component plasma (OCP), where the charges interact with a log potential and move on the curved surface of a cylinder have been investigated numerically. For vacancy–interstitial pairs, the log term of the direct Coulomb attraction and the elastic displacement energy cancel exactly at long distances, resulting in a defect energy of $O(1)$. The numerical results confirm the predicted asymptotic behavior but also identify critical distances below which pairs evolve to different forms. We have found that bound pairs of dislocations --- created by adding/removing 120 degree "zig-zags" of particles --- have a dependence on their preparation history which is not accounted for in the usual starting point of the KTHNY theory.   Furthermore, isolated  dislocations, whose presence   disrupts crystalline order, have  energies  of $O(1)$ at some values of $N$, the number of particles in the system, and therefore will be thermally excited, raising questions about the applicability of standard KTHNY theory to the OCP, and supporting older suggestions that there are no phase transitions at all in the two-dimensional OCP.

\end{abstract}

\maketitle

\section{Introduction}
\label{sec:intro}

We study in this paper the energetics and nature of  defects in the two-dimensional (2D) one component plasma (OCP) from its ground state. This model is encountered in a variety of contexts, ranging from the quantum Hall effect to dusty plasmas. Charges are confined to move on a two-dimensional surface and repel each other via the long-range 2D Coulomb potential, which varies with distance $r$ as $-\ln r$. One of the reasons for studying the 2D OCP is its relation to vortices (flux lines) in the lowest Landau level approximation in thin superconducting films. These also repel each other with a logarithmic potential \cite{Tesanovic:91, ONeill:93}. There have been numerous simulational studies of both these systems over many years, see e.g. \cite{Cardoso,Choquard:83, MooreGarrido:99, ONeill:93, Kato:93, MacDonald:93, Stroud:94, Dodgson:97, Walet:13}. Because of the long-range of the interactions between the particles (or flux lines for superconducting films) only rather small systems can be studied, leaving doubts as to whether these systems have the two phase transitions predicted by KTHNY theory \cite{Kosterlitz:73, Halperin:79, Young:79}. On this scenario as the temperature is lowered from high temperatures the first transition from a liquid state is to a phase with hexatic order only, while further reduction of the temperature leads to a solid state with a finite shear modulus. KTHNY theory attributes the melting of the solid state to the unbinding of bound pairs of dislocations of opposite Burger's vectors, while the hexatic phase changes to the liquid state when  bound pairs of disclinations unbind.

We have studied numerically the energy cost of creating pairs of anti-parallel dislocations for the OCP where the charges move on the curved surface of a cylinder. We have found that the positions of the two dislocations and their energy is not that predicted by   elasticity theory \cite{AlastueyJancovici}. In conventional elasticity theory the energy depends  only on the positions and Burgers vectors of the dislocations but we have found that it also depends  on how the dislocations are introduced into the system. We believe that this is a consequence of the long-range interactions in the OCP, but we have been unable to provide a quantitative theory of how it arises. For vacancies and interstitials, however, we have been able to account for the effects of the logarithmic long-range interaction explicitly.

  An upper bound on the temperature at which unbinding of anti-parallel dislocation pairs can occur is provided by the Thouless criterion \cite
{Thouless:78}. He argued that isolated dislocations would arise when their free energy turned negative. Then the solid phase would melt to a liquid. The energy of an isolated dislocation $E_d$ is given by

\begin{equation}
 E_d= \dfrac{b^2}{2 \pi} \dfrac{\mu( \lambda +\mu)} {\lambda + 2 \mu} \ln (L/a),   
\end{equation}
for a dislocation whose Burger's vectors is $\mathbf{b}$.  The Lam\'{e} coefficient $\mu$ is related to the shear modulus, and $\lambda$ is the Lam\'{e} coefficient related to compressions.  $L$  is the linear dimension of the system and $a$ is the core size of the dislocation which is of the order of the lattice spacing. The entropy $S_d$ of the isolated dislocation goes as $\approx \ln [(L/a)^2]$ as the dislocation could be near any of the sites of the lattice, so the free energy goes to zero at the temperature $T_{max}$ when the free energy $E_d -T S_d$ equals zero, so 

\begin{equation}
T_{max} = \dfrac{b^2}{4 \pi} \dfrac{\mu( \lambda +\mu) } {\lambda + 2 \mu}. 
\label{eqn:Tmax}
\end{equation} 
The size of the system has dropped out of this estimate of the melting temperature. It is an upper bound as the shear modulus $\mu$ is renormalized down from its zero temperature value by the thermal activation of bound pairs of opposite dislocations, as such bound pairs have an energy of $O(1)$. The zero-temperature values of $\lambda$ and $\mu$ can be calculated and it is these which are used in the Thouless bound.  As isolated dislocations can move freely when stresses are applied to the system, their presence signals the system will behave as a liquid. In KTHNY theory there are only bound pairs of dislocations at finite temperatures within the solid phase.

 It is conventional in OCP studies to set the density $n$ of the system to be unity. If the number of charges is $N$ in an area $A$, then $n= N/A =1$. The Burger's vector $b$ is just the lattice constant $a_0$ of the triangular lattice which forms in the ground state and so $n =  1/(\sqrt{3} b^2/2)$, and $b^2 = 2/\sqrt{3}$. For the 2D log potential $-e^2 \ln r$, Alastuey and Jancovici   \cite{AlastueyJancovici} showed that $\mu=e^2/8$ while $\lambda$ was infinite as the OCP is essentially incompressible. It then follows \cite{Khrapak:16} that 
\begin{equation}
\Gamma \equiv e^2/T_{max} = 16 \sqrt{3}\pi \approx 87.
\label{eqn:Gamma}
\end{equation}
Some simulational studies e.g. \cite{Choquard:83,Cardoso} report that there appeared to be a single \lq \lq weakly first order transition" at a value of $\Gamma \approx 140$, a result which is consistent with the Thouless bound. A similar first-order transition for the melting of the Abrikosov lattice in a superconducting film was found in Ref. \cite{Kato:93, MacDonald:93}. In Ref. \cite{MooreGarrido:99} no evidence was found for any transition in the OCP, for charges confined to the surface of a sphere. It was argued that in the liquid phase disclinations cost very little energy and an estimate of its form provided a good fit to the growing  order as the temperature is reduced towards zero. A similar situation was found in Ref. \cite{Dodgson:97} for superconducting vortices on the surface of a sphere. The growing correlations fit well to analytical approximations based on the parquet approximation \cite{YeoMoore:96, YeoPark:06} which also predicts that in the vortex liquid that there should be no phase transition, but just increasing amounts of crystalline order as the temperature is lowered towards zero. On the other hand simulations performed with periodic boundary conditions for superconducting vortices find evidence which appears to be more consistent with the two phase transition scenario of KTHNY \cite{Stroud:94,Walet:13}.

The variety of results which have been obtained from simulations are probably just a consequence of finite size effects. It is unlikely that there will be any development which will alter this situation radically in the foreseeable future and allow large systems to be studied. As a consequence, in this paper we have adopted a completely different approach, which is to examine the nature of the low-lying stationary points (in particular, local minima) above the ground state of the system. These are just the defects of the crystalline ground state, such as vacancies, interstitials and dislocations.  
In our work the two-dimensional surface is that of the curved surface of a cylinder. The OCP with this geometry have been recently studied in some detail by Cardoso et al. \cite{Cardoso_2021,Cardoso}. 
Topologically  this surface is equivalent to a flat surface (i.e. its Euler characteristic $\chi = 0$); (the surface of a sphere has Euler characteristic of 2 which  produces  an excess of  12 disclinations --- rings with 5 sides. This makes finding the ground state challenging \cite{MooreGarrido:99}). However, on the surface of the cylinder the task is much easier and for some values of $N$, where $N$ is the number of particles, it is easy to do numerically. In particular when $N$ can be written as $N = m^2$, the arrangement of the particles is close to that of a triangular lattice with no defects present (see Fig. 1(a)), although there are some small distortions from a perfect triangular lattice which are detectable at the edges of the system \cite{Cardoso_2021, Cardoso}. When $N$ is not of the form $m^2$ the ground state usually contains defects of one kind or another.

It has been observed that there are a large number of low-lying minima to be found \cite{Walet:13} in these systems.  We have found that when $N = m^2-(m-1)/2$  the ground state contains two anti-parallel dislocations and there are $m$-fold  degenerate excited states above it each  with an isolated dislocation, at a (small) energy of $O(1)$  above that of the ground state. Hence by the Thouless criterion, at any non-zero temperature,  isolated dislocations will be present in such a system, and the system would behave as a liquid. A similar result is found when $N= m^2-(m-3)/2$. What we have  been unable to show is that there are defects of finite energy which can induce the liquid state to exist at all temperatures for \textit{any} value of $N$. But our work suggests that at the very least KTHNY theory would need modification before it can be applied to the OCP. For example we have found evidence (see Sec. \ref{sec:120discs}) that two anti-parallel dislocations with their Burger's vectors at $120^{\circ}$ to the $x$-axis have their minimum energy in positions which suggest that there might be an effective energy cost for making local rotations. In the presence of such a potential, which breaks rotational invariance,  Halperin and Nelson  \cite{Halperin:79} argued that a hexatic phase transition would not exist.

In Sec. \ref{sec:OCP} we introduce the mathematical form of the OCP on the surface of the cylinder.  In Sec. \ref{sec:VacancyInters} we examine vacancy and interstitial pairs both analytically and numerically for the OCP. In Sec. \ref{sec:discolations} we describe our numerical studies of pairs of anti-parallel dislocations, and show that for the OCP the results cannot be explained using the conventional starting point of KTHNY theories.

\section{The OCP in a cylindrical geometry}
\label{sec:OCP}
In our study the charged particles move on the curved surface of a cylinder. Following the work of Cardoso et al. \cite{Cardoso_2021, Cardoso} the potential energy function to be minimized is
\begin{equation}
     E = e^2\sum_{1 \leq j < k \leq N} v(\mathbf{x}_j, \mathbf{x}_k) + e^2 \sum_{j=1}^N W(\mathbf{x}_j)
\end{equation}
where the logarithmic interaction potential and harmonic confining potential for a cylinder of radius $R$ are
\begin{equation}
     v(\mathbf{x},\mathbf{x}^{\prime}) = -\ln \left[ 2\cosh \left( \frac{x-x^{\prime}}{R} \right) - 2\cos \left( \frac{y-y^{\prime}}{R} \right) \right]^{1/2},
\end{equation}
and
\begin{equation}
     W(\mathbf{x}) = \pi n x^2,
\end{equation}
respectively, where $n$ is the areal density of the particles. We shall set its value to unity, which fixes the length scale in our problem.  Here $\mathbf{x} = (x, y)$, $y$ parametrizes the circumference, with $y \sim y + 2\pi R$. In these coordinates the metric is $ds^2= dx^2 +dy^2$, and the Laplace-Beltrami operator is $\Delta= (\partial/\partial x)^2+ (\partial/\partial y)^2$. For the Figures in this paper we have rolled the surface of the cylinder out into a flat plane, with the $y$ axis being the direction of the circumference of the cylinder. Because the charged particles repel each other a confining potential like $w(x)$ is needed to trap them into a finite region. The chosen form makes for an OCP of nearly constant density.

 The area of the unit cell of the triangular lattice, with lattice parameter $a_0$, is $\sqrt{3}/2 a_0^2$, so if we set the areal density $n=1$ then
\begin{equation}
  n=1 = 1/(\sqrt{3} a_0^2/2).
\end{equation}
The cylinder radius $R$ for a triangular lattice with $N = m^2$ particles
\begin{equation}
2 \pi R = m a_0,
\end{equation}
so that exactly $m$ charges, separated by $a_0$, fit along the circumference. The particles are very close to being in a perfect triangular lattice with $m$ equally spaced lattice planes occupying the region $x \in [-x_0, x_0]$, where
\begin{equation}
x_0 = \frac{(m-1)d_0}{2}, \qquad d_0 = \frac{\sqrt{3}a_0}{2}.
\end{equation}
The small deviations from a perfect triangular lattice occur near the edges of the confined regions and have been studied in some detail in Refs. \cite{Cardoso_2021, Cardoso}.

The potential $v(\mathbf{x},\mathbf{x}^{\prime})$ is periodic in the $y$-direction. If both $(x-x^{\prime})/R$ and $(y-y^{\prime})/R$ are both small compared to $1$, it reduces to the 2D form of the Coulomb potential in flat space $-\ln (|\mathbf{x}-\mathbf{x^{\prime}}|/R)$.  

All numerical work was performed with Mathematica's inbuilt FindMinimum function using conjugate gradients, which requires making an initial guess for the particle positions.

\begin{figure}[h]
  \centering

  % ---------------------------
  % Panel (a): full width with spyglasses
  % ---------------------------
  \begin{tikzpicture}[spy using outlines={rectangle, magnification=2, connect spies, size=40,draw=black}]
    \node (main) at (0,0) {\includegraphics[width=\linewidth]{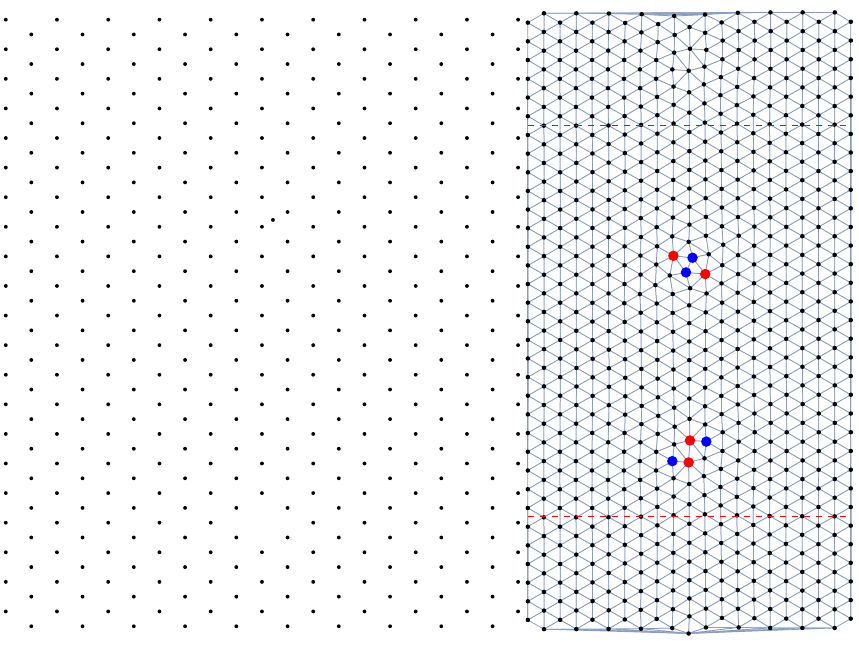}};
    \spy on (-1.7,1) in node [fill=gray!10] at (0,2);
    \spy on (-1.7,-2) in node [fill=gray!10] at (-3.3,0);
    \node[anchor=north west, font=\bfseries, xshift=1.6mm, yshift=-2mm] at (main.north west) {(a)};
  \end{tikzpicture}

  \vspace{0.25 cm}

  % ---------------------------
  % Panel (b): full width
  % ---------------------------
  \begin{tikzpicture}
    \node (b) {\includegraphics[width=\linewidth]{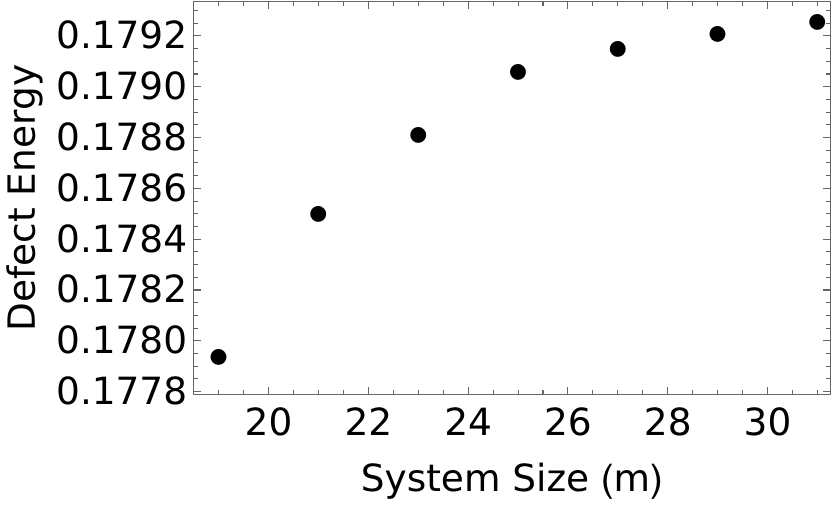}};
    \node[anchor=north west, font=\bfseries, xshift=10mm, yshift=-0.5mm] at (b.north west) {(b)};
  \end{tikzpicture}

  \caption{Initial positions, Delaunay triangulations and defect energies for vacancies and interstitials. (a) Initial positions and Delaunay triangulations for a vacancy-interstitial pair separated in the y direction. A particle on the y axis is moved slightly to the right and $(m-1)a_0/2$ up the y axis. The red dashed lines indicate the periodic boundary conditions. Red and blue dots represent 7- and 5-fold disclinations, respectively. (b) Defect energy for the vacancy-interstitial pair separated vertically as a function of $m$. when the separation $s= (m-1) a_0/2$.  This energy seems to approach a finite value as $m$ becomes large, as was expected from our analytical calculation.}
  \label{fig:vi-combined}
\end{figure}

\begin{figure}[h]
\centering

% ---------- top row ----------
\begin{minipage}[t]{0.48\linewidth}
  \centering
  \includegraphics[width=\linewidth]{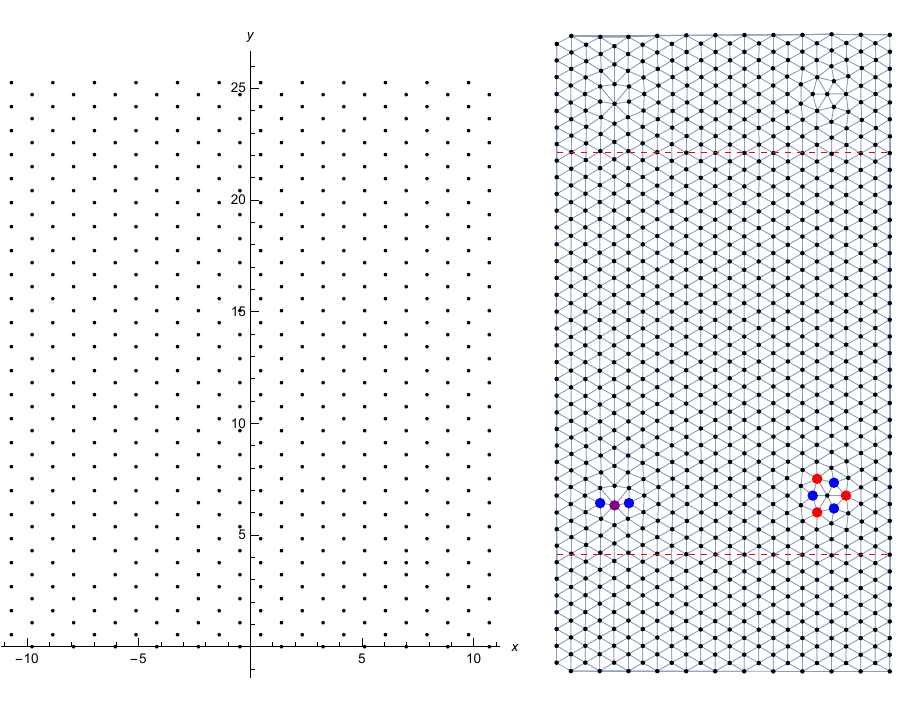}
  \textbf{(a)}
\end{minipage}
\hfill
\begin{minipage}[t]{0.48\linewidth}
  \centering
  \includegraphics[width=\linewidth]{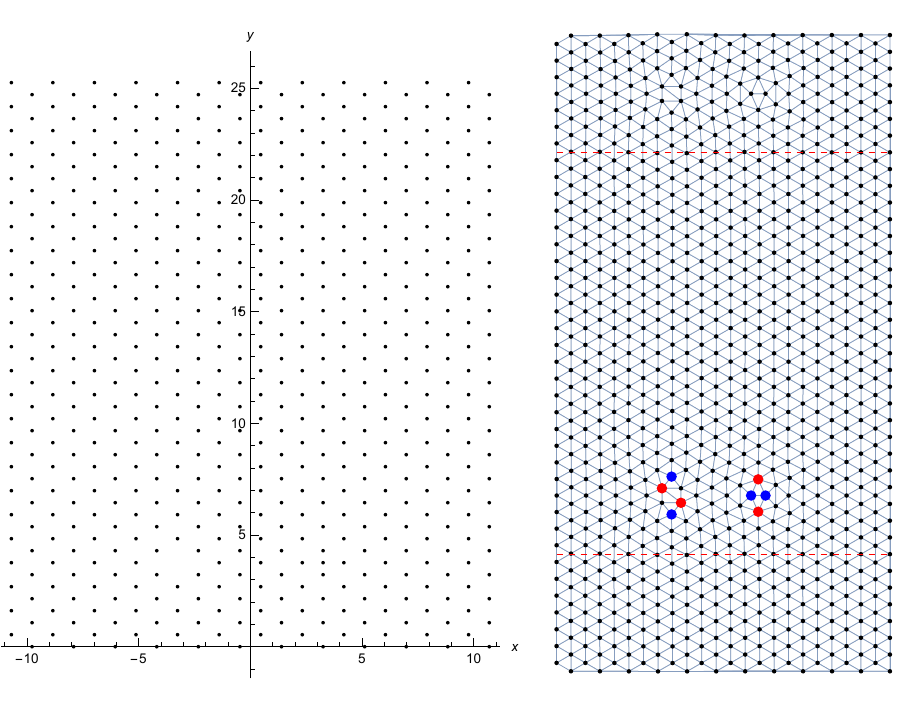}
  \textbf{(b)}
\end{minipage}

\vspace{0.4cm}

% ---------- bottom row ----------
\begin{minipage}[t]{0.48\linewidth}
  \centering
  \includegraphics[width=\linewidth]{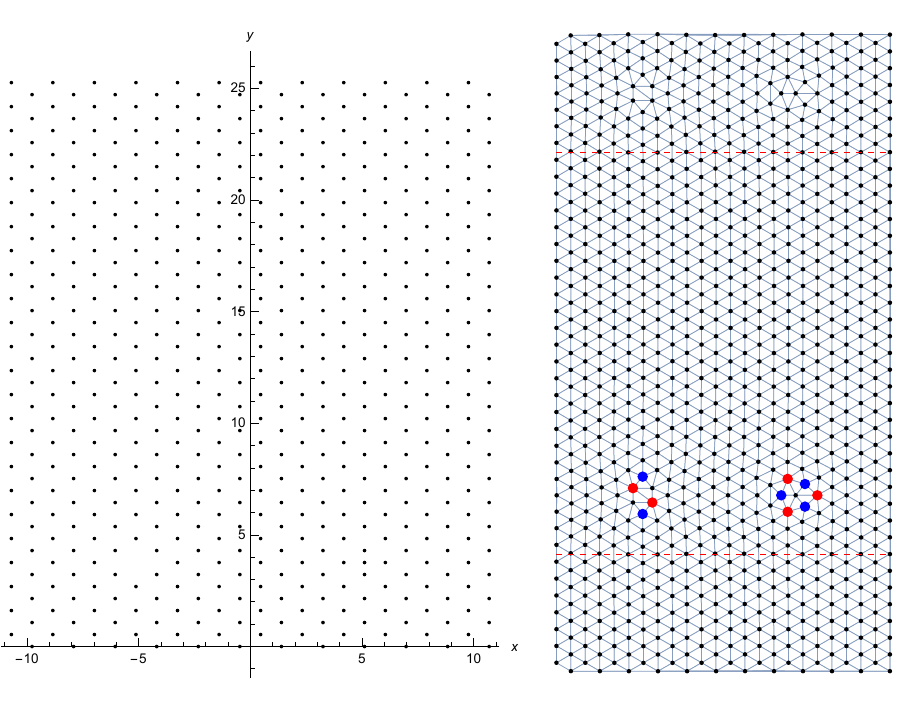}
  \textbf{(c)}
\end{minipage}
\hfill
\begin{minipage}[t]{0.48\linewidth}
  \centering
  \includegraphics[width=\linewidth]{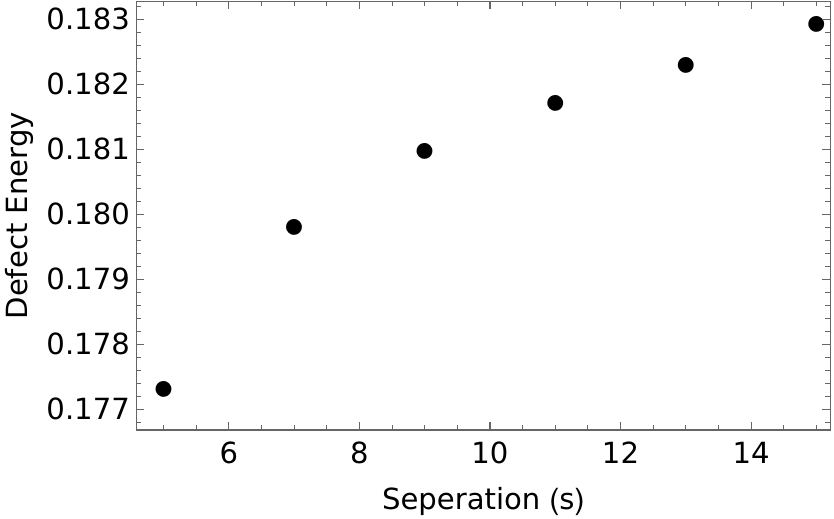}
  \textbf{(d)}
\end{minipage}

\caption{
(a) Vacancy of type $SV$ and interstitial of type $I_3$ for $m=24$, $s = 15$. The purple dot indicates that the particle has eight neighbours.
(b) Vacancy of type $V_2$ and interstitial $I_2$ for $m=24$, $s=5$. 
(c) Vacancy of type $V_2$ and interstitial of type $I_3$ for $m=24$, $s = 11$. 
(d) Defect energy for a vacancy–interstitial pair separated in the x direction as a function of $s$. 
}
\label{fig:combined}
\end{figure}

\section{Vacancies and Interstitials}
\label{sec:VacancyInters}

\subsection{Vacancy Interstitial Attraction Screening}
The direct Coulomb attraction between a centered interstitial and a vacancy with effective charges $+e$ and $-e$ respectively, separated by a distance $s$ on the cylinder is
\begin{equation}
    -e^2 \ln{\frac{s}{R}},
    \label{CoulombEnergy}
\end{equation}
provided $ s << R$.  There is also a contribution to the energy of the interstitial-vacancy pair  from the elastic energy  due to the particles moving towards the vacancy  and away from the interstitial. We show next that this energy scales as
\begin{equation}
+ e^2 \ln{\frac{s}{R}} + \mathcal{O}(1),
\label{ElasticEnergy}
\end{equation}
which precisely cancels the logarithmic dependence of the coulomb attraction and leaves an energy cost of $\mathcal{O}(1)$ to create the pair. To demonstrate this cancellation explicitly, we consider the displacement fields induced by the vacancy and interstitial in the lattice. Following Fisher et al. \cite{Fisher} for the vacancy,  there will be displacements at sites $l$ of
\begin{equation}
      u_{\alpha} (l)\sim -\frac{A_c}{2 \pi} \frac{x_{\alpha}(l)}{|\vec{x}(l)|^2}.
      \label{Dislacements}
\end{equation}
Here $A_c$ is related to $n$ via $A_c = 1/n$. The Fourier transform of Eq. (\eqref{Dislacements}), defined as
\begin{equation}
    u_{\alpha}(\vec{q})= \sum_l e^{i \vec{q} \cdot \vec{d}_l}
 u_{\alpha}(l),
 \label{FourierTransform}
\end{equation}
is
\begin{equation}
     u_{\alpha}(\vec{q})=+ i \frac{q_{\alpha}}{|\vec{q}|^2},
\end{equation}
when making the approximation $\sum_l\to  \int d^2 r /A_c$. The interstitial follows a similar expression but of differing sign. Considering both the vacancy and interstitial separated by a distance $s$ we get
\begin{equation}
    u_{\alpha}(\vec{q})= \frac{i q_{\alpha}}{|\vec{q}|^2}(1-e^{ i \vec{q}\cdot \vec{s}}).
\end{equation}
According to Alastuey and Jancovici \cite{AlastueyJancovici}, the energy elastic energy associated by these displacements is given by
\begin{equation}
    \delta E(s) = \frac{1}{2}\, \sum_q \, u_{\alpha}(\vec{q}) D_{\alpha \beta} (\vec{q}) u_{\beta}(-\vec{q}),
\end{equation}
where the dynamical matrix in the long wavelength limit is
\begin{equation}
    D_{\alpha \beta}(\vec{q})=2 \pi n e^2\frac{q_{\alpha} q_{\beta}}{|\vec{q}|^2}+\frac{e^2}{8}(q^2 \delta_{\alpha \beta}- 2 q_{\alpha} q_{\beta}).
    \label{eq:Hessian}
\end{equation}
There are corrections to this form for $ D_{\alpha \beta}(\vec{q})$ outside the long-wavelength limit. $\Sigma_q$ is shorthand for $A_c \int d^2 q  /(2 \pi)^2$, where the integral is over the first Brillouin zone of the triangular lattice. Note that the integral over the Brillouin zone can be approximated as
\begin{equation}
    A_c \int_{\text{BZ}} \frac{d^2 q}{(2 \pi)^2} \approx A_c \int_0^{k_D} \frac{2 \pi q \, dq}{(2 \pi)^2} = 1.
\end{equation}
From this,
\begin{equation}
    k_D = \sqrt{\frac{4 \pi}{A_c}}.
\end{equation}
Substituting the $u_{\alpha}(\vec{q})$ into the for $\delta E(R)$ we get
\begin{equation}
    \delta E(s)\approx e^2\int_0 ^{k_D} dq \frac{1-J_{0}(qs)}{q},
\end{equation}
where we have approximated the integral over the Brillouin zone by the integral over a circle of radius $k_D$. The integral is divergent at its lower limit, and has to be cut-off at $q$ values of $O(1/R)$. The upper limit can be replaced by $\infty$. The final result is that the elastic energy contributions go as $e^2 \ln (s/R) +O(1)$. Thus, the energy cost of the small displacements of the lattice exactly cancel the $e^2 \ln s/R$ of the direct Coulomb interaction between the vacancy and the interstitial dependence out of the vacancy-interstitial energy, leaving it of $\mathcal{O}(1)$. We next see the extent to which numerical work confirms these arguments.

\subsection{Numerical Work}

Following the analytical treatment, we now turn to numerical methods to validate these theoretical expectations, and examine the behavior of the system.
We begin by studying how the defect energy varies as a function of the distance between the vacancy and interstitial $s$. The vacancies and interstitials  have been studied near the central column, see Fig.\ref{fig:vi-combined}a. To compare the defect energies for varying $s$ we choose to move the particle roughly $(m-1)a_0/2$ up the y axis and a slight displacement in the positive x direction (for odd values of $m$). The data in Fig.\ref{fig:vi-combined}b confirm that the defect energy follows the asymptotic behavior as $m \to \infty$, as predicted by the theoretical calculation; the final data points differ by a small amount of $10^{-5}$.  The Delaunay triangulation allows us to see the form  of the dislocations associated with the vacancy and the interstitial. This can change as a function of $m$ and $s$.  See Figs. 2-4 for various kinds of vacancy-interstitial pairs along the $x$-axis.  As stated, there exists stable vacancy-interstitial pairs for different values of separation and they have been labelled by the classification in \cite{BaojiHe2013}. All forms have been observed in our system, with the exception of $I_4$, for both the vertical and horizontal separations. An analogous form for the interstitial of the SV vacancy has also been seen, where there are two particles with 7 neighbours and one particle with 4 neighbours in the centre. This was not discussed in \cite{BaojiHe2013}. One can see that the defect energy is tending to a constant value as with the vertical case for large $m$ and $s$. In reality there is a dip in the defect energy when $s$ gets large enough for the horizontal case, but this effect occurs at larger values of $s$ with increasing $m$. Therefore this is most likely nothing but a finite system size effect when the vacancies and interstitials get close to the edge.

Because the energy cost of vacancy-interstitials pairs is $O(1)$, thermal fluctuations will produce them at all temperatures, although their numbers will naturally reduce as the temperature is lowered towards zero. In Ref. \cite{Cardoso} snapshots of the particle positions in Monte Carlo simulations were produced at various temperatures and bound pairs of dislocations as in Fig. 1 are clearly visible. These correspond to vacancies or interstitials.

\section{Dislocations}
\label{sec:discolations}
\begin{figure*}[htb]
    \centering
    \includegraphics[width=0.48\textwidth]{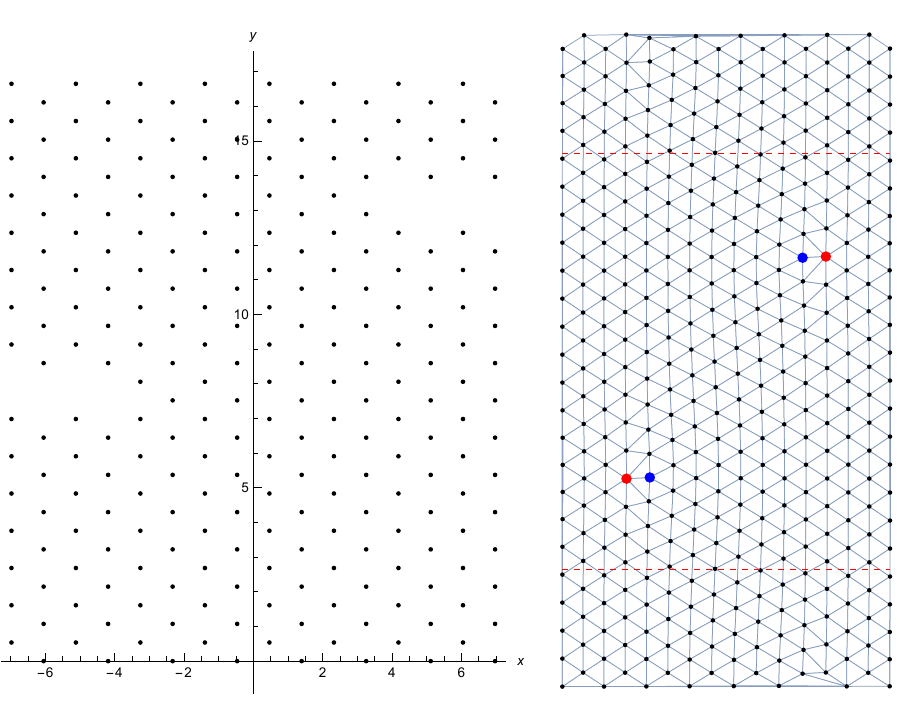}
    \hfill
    \includegraphics[width=0.48\textwidth]{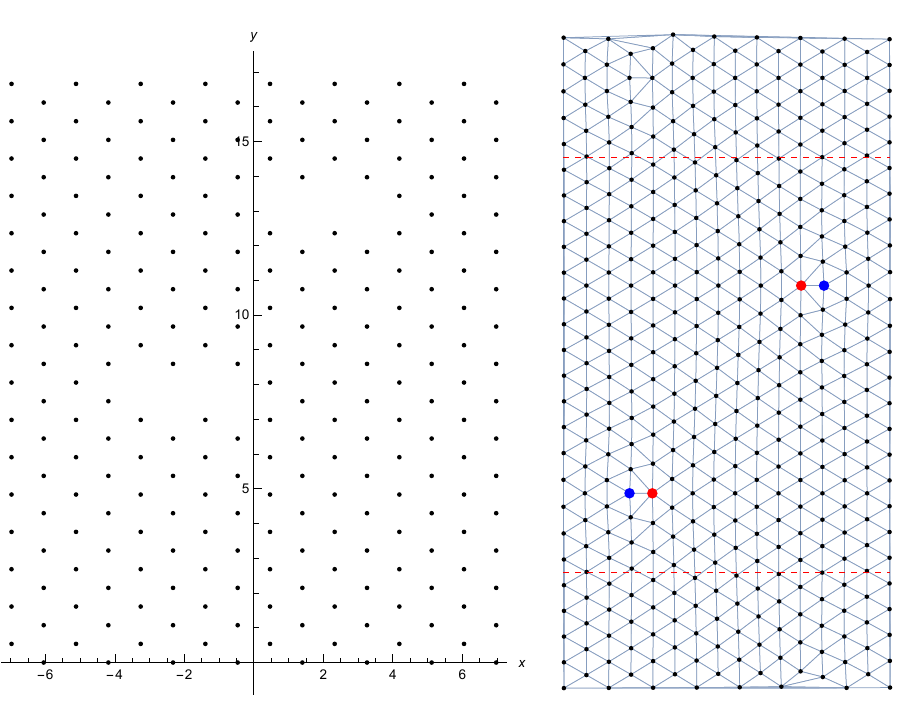}
    \caption{Two similar states with the only difference in their Delaunay triangulations being the order of the 7-5 pairs disclinations (or the direction of the Burger's vector $\mathbf{b}$). Their energies differ by $O(10^{-2})$. The  method depicted on the left figure  contains 4 columns near both edges with $m-1$ particles in both the initial and final state, while the method depicted in the right figure results in the 8 central columns having $m-1$ particles in both the initial and final state after the minimization. This state has the lower energy.  }
    \label{fig:TwoMethods}
\end{figure*}

\begin{figure}[htb]
  \includegraphics[width=0.48\textwidth]{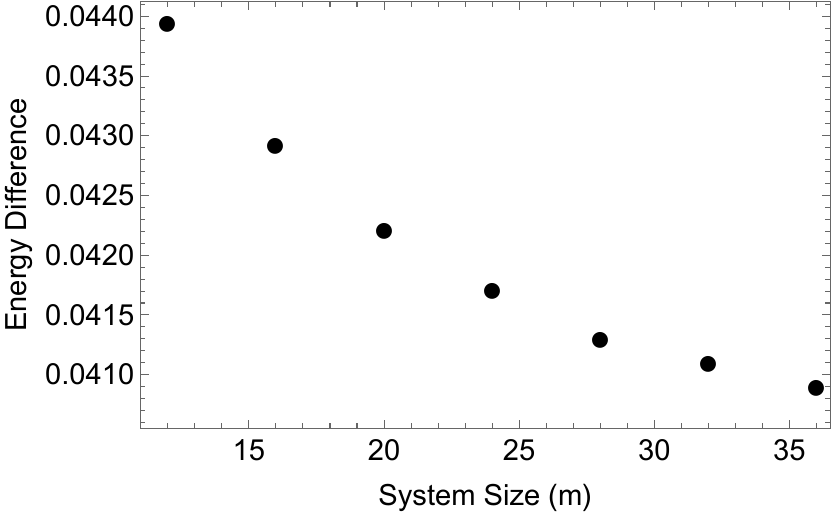}
  \caption{The difference in energy between the two almost identical states of Fig.\ref{fig:TwoMethods} as a function of $m$. The state of lower energy is the state in which the central particles are removed.  }
  \label{fig:SameStateAnalysis}
\end{figure}

We move on from vacancies and interstitials to dislocations. Alas, for them we are unable to provide any analytic treatment for how the standard calculations have to be modified in the OCP, so our work is based entirely on numerical studies, and the features induced by different methods of introducing dislocations into the system.

The energy of two dislocations with Burger's vectors $\mathbf{b}_1$ at $\mathbf{r}_1$ and $\mathbf{b}_2$ at $\mathbf{r}_2$ on the flat plane is \cite{Halperin:79}
\begin{align}
 E  = -&\frac{K_1}{8 \pi} \mathbf{b}_1 \cdot \mathbf{b}_2 \,\ln \left( \frac{|\mathbf{r}_1 - \mathbf{r}_2|}{a} \right) \notag \\
 +& \frac{K_2}{8 \pi} \frac{\mathbf{b}_1 \cdot (\mathbf{r}_1-\mathbf{r}_2)\, \mathbf{b}_2 \cdot(\mathbf{r}_1-\mathbf{r}_2)}{|\mathbf{r}_1-\mathbf{r}_2|^2} \notag\\
 +& E_c(|\mathbf{b}_1|^2+|\mathbf{b}_2|^2), 
 \label{eq:energy2disc}
\end{align}
where $a$ is a length scale related to the core size of the dislocation, $E_c$ is the core energy of a dislocation, and $K_1$ and $K_2$ are related to the Lam\'{e} coefficients $\lambda$, $\mu$ and $\gamma$ of the elasticity expression \cite{Halperin:79}
\begin{equation}
H_E=\frac{1}{2 a_0^2}\int d^2 r[2 \mu u_{ij}^2+ \lambda u_{kk}^2 + \gamma(\partial _y u_x-\partial_x u_y)^2],
\label{eq: defelast}
\end{equation}
where 
\begin{equation}
  u_{ij}= \frac{1}{2}\left[\frac{\partial u_i(\mathbf{r})}{\partial r_j}+\frac{\partial u_j(\mathbf{r})}{\partial r_i}\right],
\end{equation}
and we have used the convention of sums over repeated indices in Eq. (\ref{eq: defelast}). $u_x$ and $u_y$ are small displacements in the $x$ and $y$ directions of the particles from the positions which they have at a minimum of the energy. The last term in Eq. (\ref{eq: defelast}) is the energy cost associated with a rotation $\theta(\mathbf{r})$, where 
\begin{equation}
\theta(\mathbf{r})=  \frac{1}{2}\left[\frac{\partial u_x}{\partial y}-\frac{\partial u_y}{\partial x}\right].   
\end{equation}
 $\gamma$ might be expected to be zero unless there are features present like an underlying substrate which lock the system into a particular orientation \cite{Halperin:79}. The coefficients $K_1$ and $K_2$ are related to the elasticity coefficients by \cite{Halperin:79}
\begin{equation}
K_1= \frac{4 \mu(\mu +\lambda)}{2 \mu+\lambda}+\frac{ 4 \mu \gamma}{\mu+\gamma},  
\end{equation}
and 
\begin{equation}
K_2= \frac{4 \mu(\mu +\lambda)}{2 \mu+\lambda}-\frac{4 \mu \gamma}{\mu+\gamma}.  
\end{equation}
Thus if $\gamma=0$, $K_1=K_2$. The work of Alastuey and Jancovici \cite{AlastueyJancovici} implies that $\gamma = 0$ at least for the OCP confined to a circular patch, and that $\mu = e^2/8$ and $\lambda + 2 \mu= 2 \pi n e^2/q^2$. The OCP is essentially incompressible, which is the origin of the divergence of $\lambda$ as the wavevector $q\to 0$.

For simplicity we shall focus mostly on the case of two anti-parallel dislocations with $\mathbf{b}_1=-\mathbf{b}_2$. On the triangular lattice there are just three possible orientations of $\mathbf{b}$, and we have studied the case $|\mathbf{b}|= b = a_0$. For situations in which $\mathbf{b}_1+\mathbf{b}_2 = 0$ i.e. anti-parallel dislocations, the core energy $E_c$ is finite and independent of the size of the system. That is, if the distance between the dislocations is $|\mathbf{s}|$ then provided $s << R$, where $R$ is the radius of the cylinder, $E_c$ should be independent of $s$ and $R$, provided also that the dislocations are not close to the edges of the cylinder, where edge effects come into play. Nelson and co-workers \cite{Nelson:13, NelsonGlide} have studied the energy of two dislocations on the surface of a cylinder and modified them to include the effects of the periodicity associated with the cylindrical symmetry and found as expected that their results reduce to those on the flat plane when $s << R$. We next examine the positions of two anti-parallel dislocations using Eq. (\ref{eq:energy2disc}).

Dislocations easily move (glide) along the direction of their Burger's vectors. It is only the Peierls-Nabarro potential (for a review see \cite{nabarrorev:97}) which opposes the gliding motion and for the OCP it seems to be very weak. In our numerical study of dislocations some initial configuration of the particles is selected and minimized by the method of conjugate gradients. We can pause the minimization before the final state is reached and observe the gliding (if any needed) to reach the final state. It is only at saddle-points that we have observed any trapping due to the Peierls-Nabarro potential.

For two anti-parallel dislocations of Burger's vectors $\mathbf{b}$ and $\mathbf{-b}$ their energy is as described by Eq. (\ref{eq:energy2disc}). If the glide directions are separated by $\mathbf{s}$,  $\mathbf{r}_1-\mathbf{r}_2 = \mathbf{s} + \alpha \mathbf{b}$, where $\mathbf{b}\cdot.\mathbf{s} = 0$, so $|\mathbf{r}_1-\mathbf{r}_2|=\sqrt{s^2 +\alpha^2 b^2}$. The energy of the two dislocations is
\begin{equation}
 E= \frac{K_1 b^2}{8 \pi} \ln \left[\frac{\sqrt{s^2 + \alpha^2 b^2}}{a}\right] -\frac{K_2}{8 \pi} \frac{\alpha^2 b^2}{s^2 + \alpha^2 b^2}. \label{eq:minimize1}  
\end{equation}
The relative spacing of the two dislocations is determined by setting the first derivative of $E$ with respect to $\alpha$ to zero,
which is 
\begin{equation}
\frac{dE}{d \alpha}=\frac{b^4 K_1 \alpha}{8 \pi (s^2+\alpha^2 b^2)^2}[\alpha^2 b^2-(2 K_2/K_1-1) s^2]= 0.
\label{eq:minimizd}
\end{equation}
If $2 K_2 \ge K_1$, the minimum is when 
\begin{equation}
\alpha^2 b^2= (2K_2/K_1-1)s^2,
\label{eq:nt}
\end{equation}
 and the solution at $\alpha = 0$ is a saddlepoint at a higher energy. When $2 K_2 < K_1$, there is only a minimum which is at $\alpha = 0$. For the solution at $\alpha = 0$ the minimum position of the two dislocations brings them to their closest positions which can be reached on their glide paths.

In the following subsections we examine the extent that Eq.(\ref{eq:energy2disc}), which is the starting point for the KTHNY theory actually applies to the OCP on the surface of a cylinder. We shall find that it fails to account for an important feature of dislocations in this system, namely that their energy is not only dependent on the positions $\mathbf{r}_i$ of the dislocations and their associated Burger's vectors $\mathbf{b}_i$, but crucially on which particles were added or removed to produce the dislocations.

The creation of dislocations in our system has exclusively consisted of removing and adding 120 degree zig-zags of particles as discussed in \cite{Fisher}. Moving/removing $K$ particles in a zig-zag to the edge of the system results in a single dislocation $K$ lattice spacings from the edge; while elsewhere leads to a pair of dislocations with opposite Burger's vectors separated by $K$ lattice spacings. These zig-zags produce a net change of one in the particle numbers in the columns in which they are placed/removed. We believe that because of the long-range of the interactions in the OCP these changes in the charge numbers can have important consequences, as will be demonstrated  in the following subsections.  

\subsection{Preparation Matters}

\begin{figure}[htb]
    \centering
    \includegraphics[width=1\linewidth]{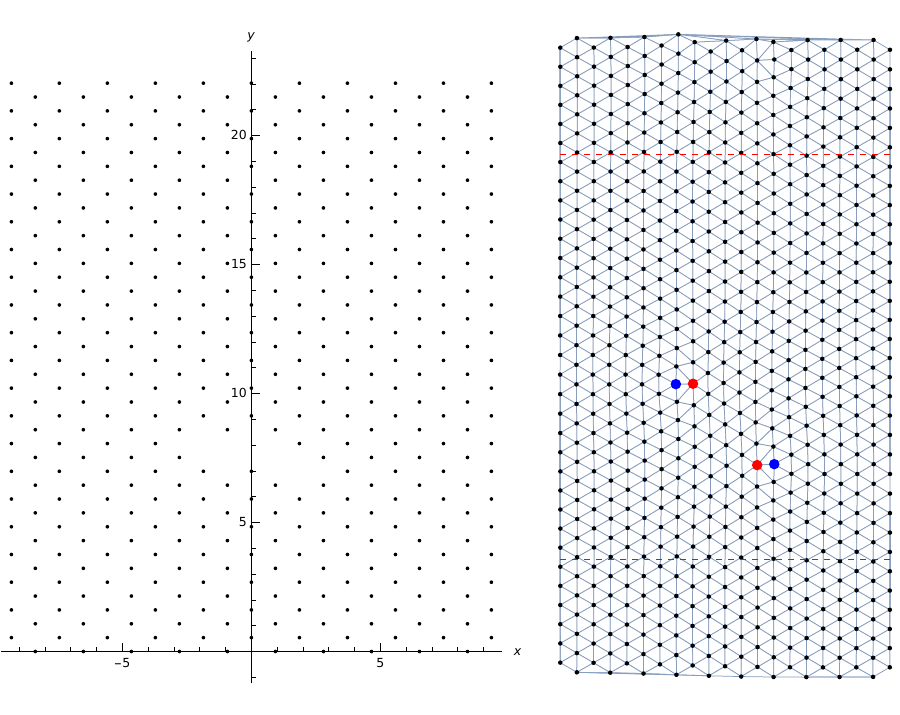}
    \caption{Two anti-parallel dislocations whose Burger's vectors are parallel to the $y$-axis in  positions consistent with the solution of Eq. (\ref{eq:nt}) i.e.  $s= K$,  where $K$ is the number of particles removed in the left figure, where $K= 5$ and with $K_1=K_2$. For larger values of $K$ the saddle solution where the dislocations lie at the same value of $y$ is found, but this solution exists only because the gliding forces at larger values of $K$ cannot overcome the Peierls-Nabarro potential. This state has a higher energy than the one depicted in the figure, differing by 0.00181. Notice that there are $K$ columns containing $m-1$ particles.  }
    \label{fig:smallerR}
\end{figure}

Moving a zig-zag of $K$ particles and minimizing generally results in $K$ columns with $m+1$ and $K$ columns with $m-1$ particles. The additional/removed charge is spread out evenly in each column. Elasticity theory insists that the preparation of state is irrelevant for the energy of the system. This seems impossible as one could create identical states in the eyes of elasticity theory, but in the column picture they would look completely different. In Fig.\ref{fig:TwoMethods} two ways of creating a state is seen. They differ in their energy by an amount  order of 10$^{-2}$ which is direct evidence for the relevance of state preparation. To ensure that this energy difference is not simply a finite-size effect, a study was conducted on the difference in energy as a function of $m$ for analogous states. The result is seen in Fig.\ref{fig:SameStateAnalysis} and the data are approaching a constant value and hence is not a finite-size effect.

The difference in the energy of these two states might be attributed to  differences induced in their core energies $E_c$ by the method of preparation. In the next two subsections we give examples where the method of preparation seems to affect the values of $K_1$ and $K_2$.

\subsection{Dislocations whose Burger's vectors are parallel to the y-axis }
\label{sec:yaxis}
 In this subsection we describe the case of two dislocations whose  vectors $\mathbf{b}$ and $-\mathbf{b}$ are parallel to the $y$-axis, as in Fig. \ref{fig:smallerR}.
Thus our results for anti-parallel dislocations with $b$-vectors parallel to the $y$-axis are entirely consistent with Eq. (\ref{eq:energy2disc}) with $K_1= K_2$, which is the starting point of KTHNY theory. However, in the next subsection we provide examples of anti-parallel dislocations which are quite incompatible with this.

\subsection{Dislocations whose Burgers vectors are at $120^{\circ}$ to the $x$-axis}
\label{sec:120discs}
\begin{figure}[htb]
    \centering
    \includegraphics[width=1\linewidth]{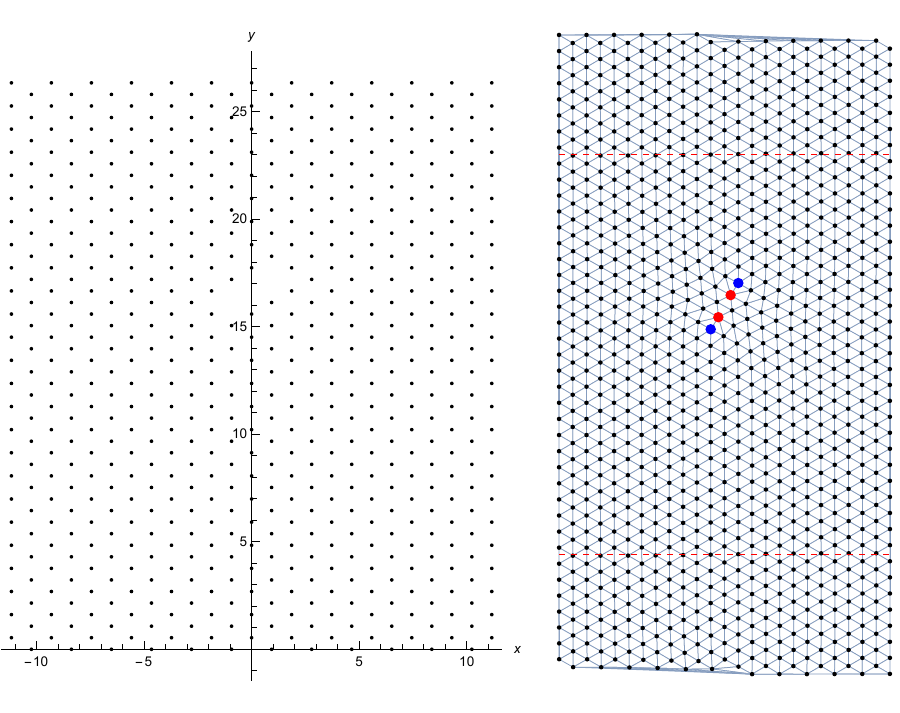}
    \caption{Two anti-parallel dislocations for $m = 25$ whose Burger's vectors are at $120^{\circ}$  to the $x$-axis in  positions consistent with a stable solution of Eq.(\ref{eq:nt}) where the dislocations are side by side on their respective glide directions, which implies $2 K_2 \le K_1$, according to Eq. (\ref{eq:minimizd}). This state has a lower energy by $0.00722$ compared to the energy of the state in Fig. \ref{fig:central120}. The state depicted has the 3 central columns missing just one charge each.}
    \label{fig:sidebyside120}
\end{figure}

\begin{figure}[htb]
    \centering
    \includegraphics[width=1\linewidth]{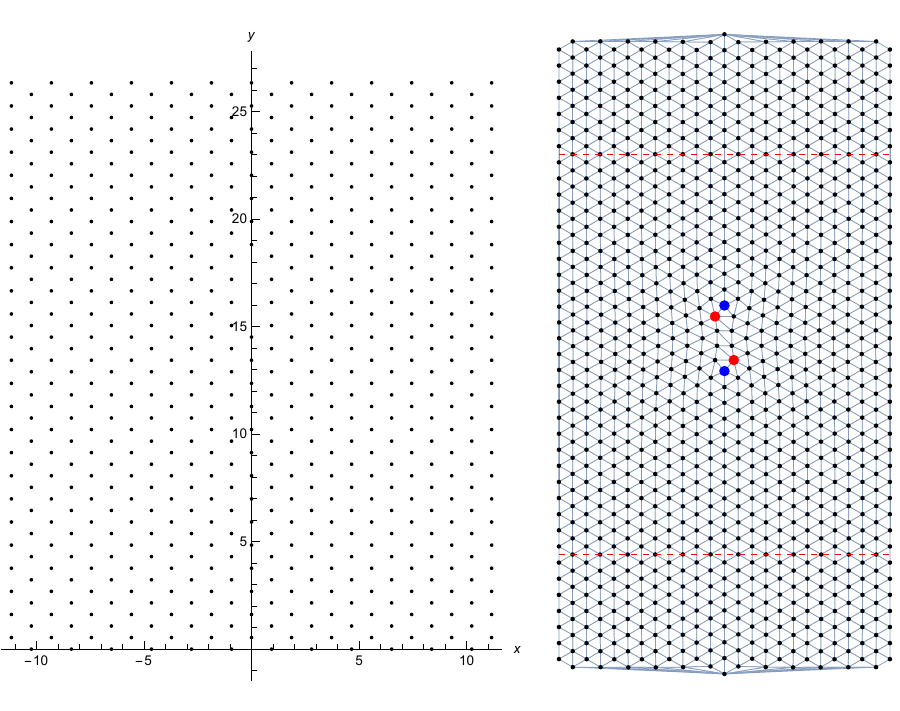}
    \caption{Two anti-parallel dislocations for $m = 25$ whose Burger's vectors are at $120^{\circ}$ to the $x$-axis.  It has a higher energy than that in Fig. \ref{fig:sidebyside120} by $0.00722$. This state has 3 particles missing from just the the central column.}
    \label{fig:central120}
\end{figure}

Figs. \ref{fig:sidebyside120} and \ref{fig:central120} provide striking illustrations of the effects of two different preparation methods.

\begin{figure}[htb]
  \centering
  \begin{tikzpicture}
    \node (a) {\includegraphics[width=0.9\linewidth]{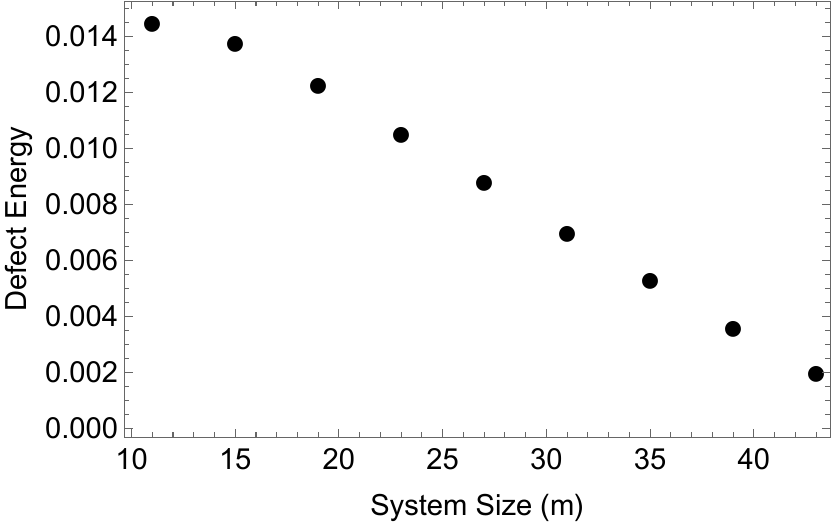}};
  \end{tikzpicture}
  \caption{The defect energy as a function of $m$ for an isolated dislocation in the center of the system. The defect energy is so small  that such isolated dislocations will be thermally excited in large systems even at low temperatures and destroy the solid state.}
  \label{fig:IsolatedDislocationEnergy}
\end{figure}

\begin{figure*}[htb]
  \centering

  \begin{minipage}{0.48\textwidth}
    \centering
    \begin{tikzpicture}
      \node[inner sep=0] (b) {\includegraphics[width=\linewidth]{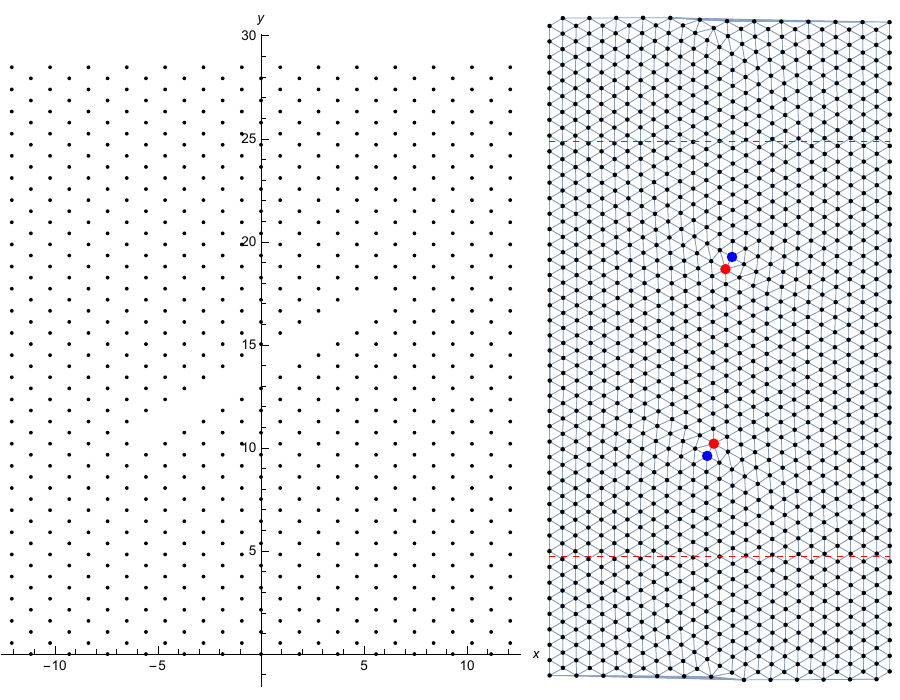}};
      \node[anchor=north west,font=\bfseries, xshift=1.6mm, yshift=-1.6mm] at (b.north west) {(a)};
    \end{tikzpicture}
  \end{minipage}%
  \hfill
  \begin{minipage}{0.48\textwidth}
    \centering
    \begin{tikzpicture}
      \node[inner sep=0] (c) {\includegraphics[width=\linewidth]{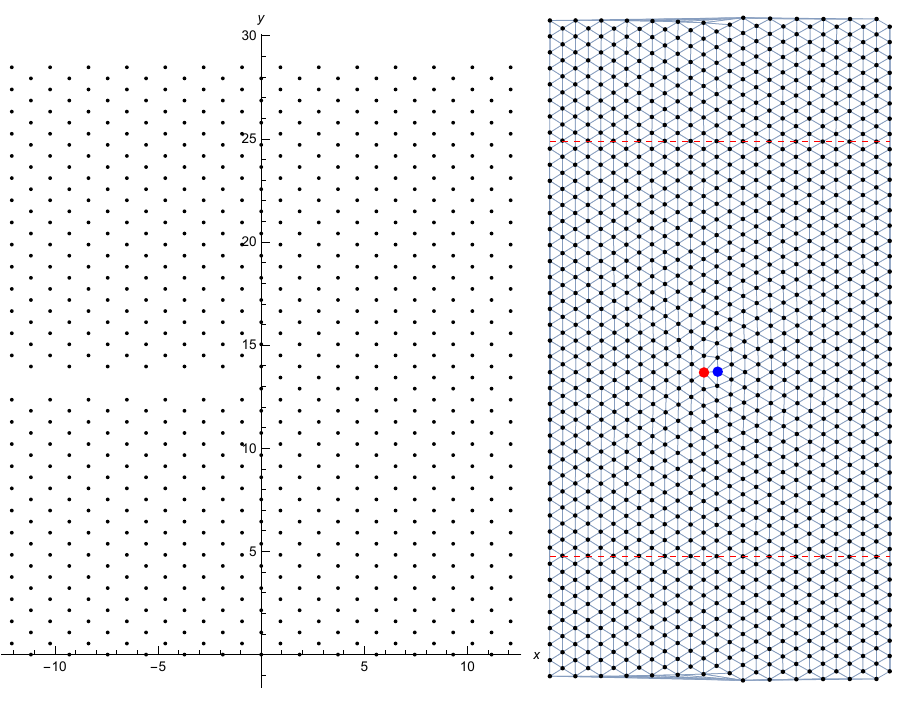}};
      \node[anchor=north west,font=\bfseries, xshift=1.6mm, yshift=-1.6mm] at (c.north west) {(b)};
    \end{tikzpicture}
    \label{fig:isolatedprep}
  \end{minipage}

  \caption{(a) The ground state for $N = m^2 -\tfrac{m-1}{2}$ particles, created by removing a symmetrical diagonal through the center of the system and minimizing. The result is the column in the center contains $m-\frac{m-1}{2}$ and all other columns contain $m$ particles. (b) An isolated dislocation in the center of the system created by removing $ \tfrac{m-1}{2}$ particles along a horizontal row to the center of the system. 
  This results in $\frac{m-1}{2}$ columns with $m-1$ particles and $\frac{m+1}{2}$ columns with $m$ particles. Both states contain the same number of particles $N$.}
  \label{fig:IsolatedDislocationConfigs}
\end{figure*}

\begin{figure}[htb]
  \includegraphics[width=0.48\textwidth]{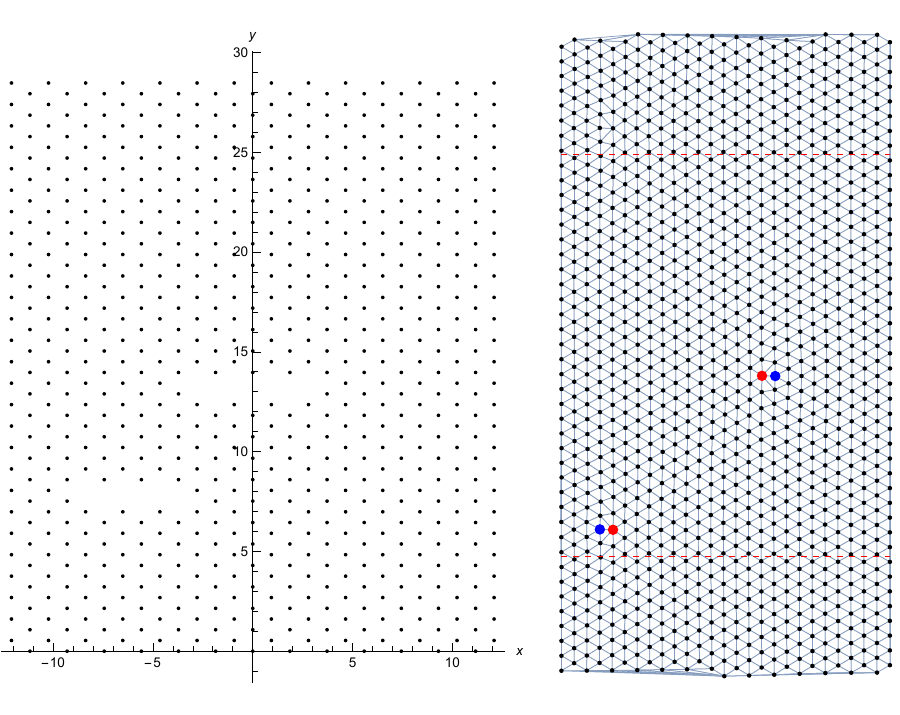}
  \caption{Method for creating pairs of anti-parallel Burgers vector dislocations separated by a distance $(m-1)/2$ for $m =27$. In the example shown the left dislocation is at a distance $K= 4$ from the edge of the system.}
  \label{fig:MovingK}
\end{figure}

\begin{figure}[htb]
  \includegraphics[width=0.48\textwidth]{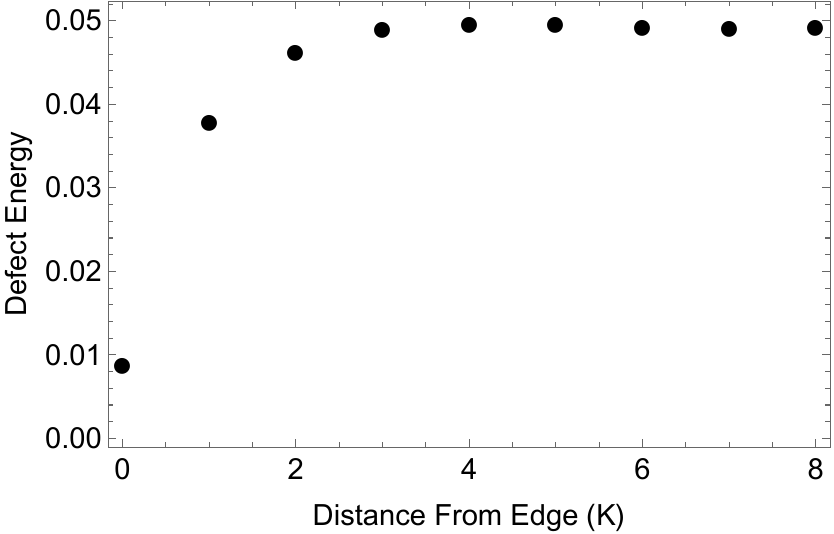}
  \caption{Energy cost of creating the pair of dislocations of Fig. \ref{fig:MovingK} as a function of $K$, the distance  of the left dislocation from the edge of the system, for $m = 27$.}
  \label{fig:energycost}
\end{figure}

Fig. \ref{fig:sidebyside120} has an initial state where one particle is removed from each of the 3 central columns, while Fig. \ref{fig:central120} has 3 particles removed from the central column. The final state of the former is lower by $0.00722$. If these states had energies given by Eq. (\ref{eq:energy2disc}) this would only be possible if $2 K_2 \le K_1$. In other words, the values of $K_1$ and $K_2$ would seem also to depend on the method of preparation of the dislocations. This of course indicates that Eq. (\ref{eq:energy2disc}) is not a satisfactory starting point for describing the energetics of dislocations in the OCP. It fails to satisfactorily account for the effects of the long-range interactions of the charged particles which are moved to make the dislocation. In addition if $K_1 \ne K_2$ then that suggests the presence of a term like $\gamma$ in the elastic energy. This would also imply an absence of rotational invariance in the system. No hexatic phase would then exist according to Ref. \cite{Halperin:79}.

\subsection{Isolated Dislocations}

Although dislocations at the edge are not enough to disrupt the crystalline order in the system as a whole, an isolated dislocation in the center would achieve this. Such a state is possible by removing $\frac{m-1}{2}$ particles in a zig-zag from the edge to the center of the system as seen in Fig.\ref{fig:IsolatedDislocationConfigs}b. To conduct a study on the energy cost of such a state the ground-state must be known, in our case for $N = m^2 - \frac{m-1}{2}$ as seen in FIG.\ref{fig:IsolatedDislocationConfigs}a. This consists of two anti-parallel dislocations in the center of the system with Burgers vectors $\pm[\frac{\sqrt{3}}{2},\frac{1}{2}]$. The defect energy of the isolated dislocation decreases as a function of $m$ as shown in Fig.\ref{fig:IsolatedDislocationEnergy}. The number of ways this dislocation can be inserted into the system is $O(m)$, so
 if one uses it with the Thouless bound, the melting temperature would apparently be at very low.

Notice that in Fig. \ref{fig:IsolatedDislocationConfigs}(a) there is an example of fractionalization \cite{Bowick:07}. In the initial state (m-1)/2 columns have each lost one particle whereas in the final state only the central column has lost particles, and it has  has lost (m-1)/2 particles. This behavior is rather unusual. Mostly the columns which were depleted or had particles added  in the initial state have the same numbers of particles after the minimization.

\subsection{Well-separated dislocations}
\label{sec:wellseparateddiscs}

Pairs of dislocations of opposite Burgers vectors can be created also at little energy cost provided that they are separated by distances comparable to the system size (see Fig. \ref{fig:MovingK}). We have studied the cases where one dislocation with its Burger's vector parallel to the y-axis is located $K$ columns from the edge of the system and the opposite anti-parallel dislocation is located (m-1)/2 columns away from it. The case $K=0$ is the case of the isolated dislocation. The initial state produces 4 dislocations but the two inner opposite dislocations are on the same glide path and annihilate each other leaving just the two opposite dislocations in the configuration expected from Sec. \ref{sec:discolations}B. Removing $(m-1)/2$ particles in a single line also produces two dislocations, but in the side-by-side position and it has a higher energy. It is stabilized by the Peierls-Nabarro potential.  In Fig. \ref{fig:energycost} we have plotted the defect energy of the states  in Fig.\ref{fig:MovingK} (above the ground state of Fig. \ref{fig:IsolatedDislocationConfigs}(a)) as a function of $K$. The isolated dislocation case $K=0$ has an unusually low energy cost, but as a function of $K$ the defect energy reaches a plateau at $0.05$. Thus there exist isolated dislocations and pairs of well-separated dislocations whose energies are finite and which will when thermally excited disrupt the crystalline order (at least for certain values of $N$).

\section{Conclusions}

For vacancy–interstitial pairs in the OCP, we have shown that the attractive logarithmic Coulomb interaction  between them is precisely canceled by the elastic energy of the small displacements they produce, leaving an O(1) defect energy which hardly varies with their separation. The energy of a dislocation has been  found to be dependent on its method of preparation. We have demonstrated that continuum elasticity misses the impact of the charges moved to create the dislocations. In particular the starting point of KTHNY theory, Eq. (\ref{eq:energy2disc}), does not describe correctly the case when the $\mathbf{b}$ vectors have a component along the axis of the cylinder, as in Sec. \ref{sec:discolations} C. We have noticed that  isolated dislocations and pairs of well-separated anti-parallel dislocations can be created at a small finite $O(1)$ energy cost from the ground state at least  for some values of $N$ at which the ground state already contains defects. This cast doubts on the existence of a melting phase transition, but is consistent with older suggestions that there are no phase transitions in the OCP \cite{MooreGarrido:99}. However, until a full theory of dislocations in the OCP allowing for the effects of the missing or added charges required to create the dislocations is available, the uncertainty will continue.

\section*{Acknowledgments}
One of us (MAM) would like to thank Prof. A. P. Young for a useful correspondence.
\nocite{*}
\bibliography{refs}

\end{document}